\documentclass{article}

\usepackage{microtype}
\usepackage{graphicx}
\usepackage{subfigure}
\usepackage{booktabs} 

\usepackage{hyperref}
\usepackage{tikz}
\usepackage{pifont}

\usepackage[accepted]{icml2025}

\usepackage{amsmath}
\usepackage{amssymb}
\usepackage{mathtools}
\usepackage{amsthm}

\usepackage[capitalize,noabbrev]{cleveref}

\theoremstyle{plain}

\theoremstyle{definition}

\theoremstyle{remark}

\usepackage[textsize=tiny]{todonotes}

\icmltitlerunning{FLAP: Fully-controllable Audio-driven Portrait Video Generation 
through 3D head conditioned diffusion model}

\begin{document}

\twocolumn[{
\icmltitle{FLAP: Fully-controllable Audio-driven Portrait Video Generation 
through 3D head conditioned diffusion model}



\icmlsetsymbol{equal}{*}

\begin{icmlauthorlist}
\icmlauthor{Lingzhou Mu}{thu}
\icmlauthor{Baiji Liu}{comp}
\icmlauthor{Ruonan Zhang}{thu}
\icmlauthor{Guiming Mo}{comp}
\icmlauthor{Jiawei Jin}{thu}
\icmlauthor{Kai Zhang}{thu}
\icmlauthor{Haozhi Huang}{comp}
\end{icmlauthorlist}

\icmlaffiliation{thu}{Tsinghua University}
\icmlaffiliation{comp}{Xverse}

\icmlcorrespondingauthor{Firstname1 Lastname1}{first1.last1@xxx.edu}
\icmlcorrespondingauthor{Firstname2 Lastname2}{first2.last2@www.uk}

\icmlkeywords{Machine Learning, ICML}

\vskip 0.3in

\renewcommand\twocolumn[1][]{#1}
\begin{center}
\vskip 0.2in
    \includegraphics[width=0.9\textwidth]{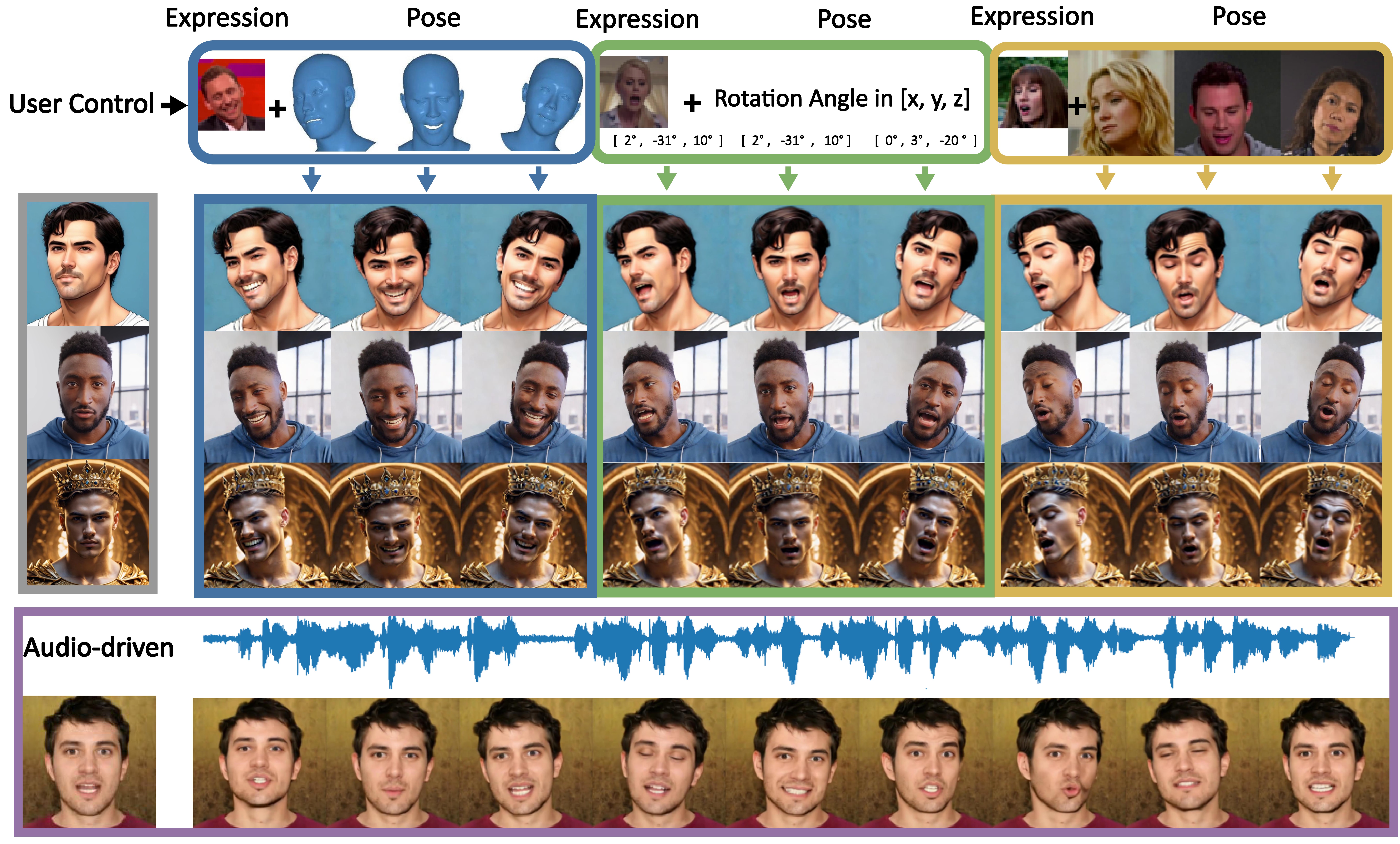}
    \begin{minipage}{0.9\textwidth}
        \centering
        \textit{Figure 1.} FLAP offers full controllability by supporting various types of control signals
    \end{minipage}
\end{center}
\vskip 0.2in
}]

\newcommand{\stars}[1]{
    \foreach \n in {1,...,#1} {
        \ding{72}
    }
}




\begin{abstract}
Diffusion-based video generation techniques have significantly improved zero-shot talking-head avatar generation, enhancing the naturalness of both head motion and facial expressions. However, existing methods suffer from poor controllability, making them less applicable to real-world scenarios such as filmmaking and live streaming for e-commerce. To address this limitation, we propose FLAP, a novel approach that integrates explicit 3D intermediate parameters (head poses and facial expressions) into the diffusion model for end-to-end generation of realistic portrait videos. The proposed architecture allows the model to generate vivid portrait videos from audio while simultaneously incorporating additional control signals, such as head rotation angles and eye-blinking frequency. Furthermore, the decoupling of head pose and facial expression allows for independent control of each, offering precise manipulation of both the avatar's pose and facial expressions. We also demonstrate its flexibility in integrating with existing 3D head generation methods, bridging the gap between 3D model-based approaches and end-to-end diffusion techniques. Extensive experiments show that our method outperforms recent audio-driven portrait video models in both naturalness and controllability.
\end{abstract}
\vspace{-1cm}

\section{Introduction}
\label{submission}
Portrait video generation, or zero-shot talking-head avatars, aims to animate a static portrait image using speech audio or video. Diffusion-based methods have demonstrated high precision in lip synchronization, realistic facial behavior, and natural head movements through specialized modules and network architectures.

However, in practical applications like live streaming for e-commerce or film production, naturalness alone is insufficient. For example, in filmmaking, an avatar should turn its head toward someone before starting a conversation. Additionally, avatars are often required to perform specific motions, such as nodding or head-shaking, as dictated by the film script.

As alternatives, LivePortrait \cite{guo2024liveportrait} and MegActor \cite{yang2024megactorsigmaunlockingflexiblemixedmodal} use video as input to generate head motions and expressions based on reference videos. These methods transform the challenge into an expression mapping task. However, by requiring additional video input to specify target expressions and motions, their usability is limited. Earlier GAN-based methods \cite{burkov2020neural,zhou2021pose} separate faces into identity and non-identity components, leveraging identity consistency and pose variation across frames. However, these methods still produce unnatural videos, with restricted pose variations and severe identity loss, resulting in faces that often appear generic. The closest attempt, VASA-1 \cite{xu2024vasa}, uses FaceVid2Vid \cite{wang2021one} and Mega-Portrait \cite{drobyshev2022megaportraits} as its motion-to-video model, while generating the motion input via a diffusion model. Although this approach enhances Mega-Portrait’s control by synthesizing controllable motion inputs, it still relies on the original Mega-Portrait for head rotation control, leading to the same generic face issue.

In contrast, more recent works use diffusion-based models to achieve superior naturalness, but at the cost of reduced controllability. For example, EchoMimic \cite{chen2024echomimic}, Ani-portrait \cite{wei2024aniportrait} and Follow-Your-Emoji \cite{ma2024follow} introduce 2D motion fields based on selected landmarks. These landmark based approaches project facial landmarks into 2D image and combines them with noisy latent variables to guide the diffusion process, resulting in three major drawbacks. 1) Projecting landmarks into 2D image space destroys the 3D information, hindering the model's ability to learn 3D structure. 2) Simply adding 2D landmark images to noisy latents destroys the translation invariance, requiring additional alignment \cite{ma2024follow,wang2024v} between the input landmarks and the reference face. 3) Facial landmarks are suboptimal as control signals for human faces, as they provide vague semantic information and are entangled with identity. Using this vague and entangled information as a primary control signal significantly limits the potential of diffusion models. Furthermore, most end-to-end diffusion-based models \cite{xu2024hallo,tian2024emo,jiang2024loopy,cui2024hallo2} do not support user-specified control signals, ignoring the importance of controllability.

To address this limitation and enhance model usability across various scenarios, we present a novel approach called FLAP. As shown in Figure 1, FLAP can generate natural head movements from audio or replace these movements with user-defined actions, such as turning, tilting, looking around, or even jerking. FLAP supports six degrees of freedom (6DoF) for control over head movements, demonstrating its nuanced understanding of 3D space. In addition to motion control, FLAP can generate lifelike expressions from three sources: audio, user-specified expression coefficients, and coefficients extracted from images or videos. Each of these sources can be used independently or combined to create new expressions. For example, as shown in when no control signal is provided, FLAP generates a talking-head video with natural expressions and mild head movements from audio alone. When expressions are provided by an image with a user-modified head pose, FLAP generates the corresponding talking-head video.

In summary, our contributions are as follows: 1) We introduce FLAP, a diffusion-based framework that produces high controllability while maintaining naturalness and high precision in lip synchronization. 2) To the best of our knowledge, FLAP is the very first work that utilizes 3D head coefficients as diffusion conditions. This not only enhanced its controllability, but also brings flexibility  in integrating with existing 3D head generation methods, bridging the gap between 3D model-based approaches and end-to-end diffusion techniques. 3) We propose a Progressively Focused Training scheme for expression-motion decoupling, in which facial expression conditions and head pose are progressively focused and learned. This ensures independent control of facial expression and head motion.

\vspace{-5pt}

\section{Related Works}

\subsection{Non-diffusion-based Portrait Animation}

For non-diffusion-based models, previous approaches \cite{zhou2020makelttalk,chen2019hierarchical} mainly introduce intermediate representation and leverage Generative Adversarial Networks (GANs) \cite{goodfellow2020generative} to generate final images. These methods first train an audio-to-motion model to predict head motion from audio, and then warp head motion into video. For example, AudioDVP \cite{wen2020photorealistic} use BFM \cite{paysan20093d} parameters as intermediate representation for motion, predicting mouth related parameters for reenactment. FaceVid2vid \cite{wang2021one} introduced 3D implicit keypoints to further enhance the representation efficiency for warping and reenactment. CodeTalker \cite{xing2023codetalker} replace BFM with FLAME \cite{li2017learning} improving expressiveness of keypoints. As for audio-to-motion models, recent works \cite{fan2022faceformer,gong2023toontalker} introduced transformer based models to enhance long sequences generalization ability and cross-modal alignment. GeneFace \cite{ye2023geneface} proposed a variational motion generator to predict landmark sequence from audio and designed a 3DMM nerf render to solve mean face problem. More recent works like Liveportrait \cite{guo2024liveportrait} introduced a mixed image-video training strategy, with a upgraded network architecture and a much larger training dataset . It achieved effectively balance of the generalization ability, computational efficiency, and controllability, yet requires video input as motion source.

\vspace{-2pt}

\subsection{Diffusion-based Portrait Animation}

Diffusion models (DMs) \cite{ho2020denoising,song2020denoising} achieves superior performance in various generative tasks including image generation \cite{rombach2022high, ruiz2023dreambooth, shi2024motion}and editing \cite{brooks2023instructpix2pix, cao2023masactrl}, video generation \cite{he2022latent, ma2024followpose,wang2024animatelcm} and editing \cite{li2024lodge,qi2023fatezero,zhang2023controlvideo}. Extensive studies \cite{zhang2023adding,li2025controlnet,hu2021lora,rombach2022high} have proven the feasibility of controlling image generation process with condition signal such as text, landmarks, segmentations, dense poses and encoded images. 

Relevant to our task of portrait animation, most recent approaches\cite{tian2024emo,chen2024echomimic,wang2024v,xu2024hallo, ma2024follow} employed a dual U-Net architecture similar to AnimateAnyone \cite{hu2024animate} and incorporate temporal module \cite{guo2023animatediff} into generation process. Besides, EMO \cite{tian2024emo} also employed face location mask and head speed as control signals using controlnet-based mechanism and attention-based mechanism respectively. Following such a paradigm, EchoMimic\cite{chen2024echomimic}, V-Express \cite{wang2024v} and Follow-your-emoji \cite{ma2024follow} integrate facial landmark conditions by adopting ControlNet \cite{zhang2023adding} mechanism. X-portrait \cite{xie2024x} proposed a ControlNet-based motion control module with local masking and scaling mechanism to achieve fine-grained motion control. Hallo2 \cite{cui2024hallo2} and Instruct Avatar \cite{wang2024instructavatar} integrate text condition as a guidance for diffusion process. VASA-1 proposed a motion space DiT \cite{peebles2023scalable} with eye gaze direction, head-to-camera distance and emotion offset as condition to generate motion latent. For motion-to-video module, it relied on a retrained FaceVid2Vid \cite{wang2021one} and MegaPortrait \cite{drobyshev2022megaportraits} to render final video result. 

Apart from studies on control signals, another branch of existing studies focus on enhancing the alignment between audio, expression and head motion. For example, Hallo \cite{xu2024hallo} proposed a hierarchical audio-driven visual synthesis module to enhance audio-visual correlation. Loopy \cite{jiang2024loopy} transformed audio feature and landmark feature into a same latent space to ensure strong dependency between audio and head motion. 

\vspace{-2pt}

\subsection{Disentanglement on Talking Head Representation}

Disentangled controlling of lip pose, expressions, and head movement is a longstanding task in talking head generation. The main target of this task is to decouple these features of a portrait video, so we can achieve independent manipulation of each. \cite{siarohin2019first} utilize explicit 2D keypoints to characterize facial dynamics. But these 2D keypoints carry insufficient head pose information in 3D space and carry identity information. Other explicit approaches like 3D face models such as BFM \cite{paysan20093d} and FLAME \cite{li2017learning} can fit head pose and expression accurately while remaining editable,  but they suffer from issues such as identity loss or lack of effective rendering methods. As solutions for expressiveness, implicit approaches such as PC-AVS \cite{zhou2021pose} employs contrastive learning to isolate the mouth space related to audio. PD-FGC \cite{wang2023progressive} extended PC-AVS by progressively disentangle identity, head pose, eye gaze direction and emotion after lip pose isolated by PC-AVS. Following this line, EDTalk \cite{tan2025edtalk} introduce a lightweight framework with orthogonal bases, and reduced training cost without performance degradation. However, the introduce of implicit features requires a preset discrete style label \cite{wang2023progressive,tan2025edtalk} or additional video as input, which reduces style diversity or cause inconvenience.

\vspace{-5pt}

\section{Preliminaries}

\subsection{Latent Diffusion Model}

Latent diffusion models (LDM) \cite{rombach2022high} are a class of diffusion models that operate diffusion process in latent space, achieving stable and fast generation. At each diffusion time step \(t\), the noisy latent \(z_t\) is fed into the network and performs cross-attention with condition \(c\). Then network predicts \(\epsilon\) conditioned on the given text or other prompts \(c\). The overall training objective can be formulated as follows:
\vspace{-5pt}
\begin{equation}
L_{LDM}=\mathbb{E}_{z_t,c,\epsilon\sim\mathcal{N}(0,1),t}\left[\|\epsilon-\epsilon_\theta(z_t,t,C)\|_2^2\right]
\label{eq:diffusion formulation}
\end{equation}
where \(\epsilon_\theta\) denotes denoising net.  In Stable Diffusion (SD), \(C\) represents text prompts $c_{text}$, image prompts $c_{img}$ or both $C=[c_{text},c_{img}]$. In the field of portrait animation, condition $C$ includes reference image, audio, context frames, landmarks and other additional information.

\vspace{-2pt}

\subsection{3D Morphable Face Model}

The research in 3D morphable face modeling aims to accurately capture, parameterize, and reconstruct human face using statistical models. As a fundamental 3D morphable face model, FLAME \cite{li2017learning} is widely adopted in numerous research works \cite{danvevcek2022emoca,danvevcek2023emotional,qian2024gaussianavatars}. FLAME use standard vertex based linear blend skinning (LBS) with blendshapes, with $N_v$= 5023 vertices, $k$ = 4 joints (neck, jaw, and eyeballs), and $N_f$ = 9976 facets. Formally, FLAME is described by a function 

\vspace{-5pt}
\begin{equation}
M(\boldsymbol{\beta},\boldsymbol{\theta},\boldsymbol{\psi})\to(\mathbf{V},\mathbf{F}),
 \label{eq:FLAME formulation}
\end{equation}

which parameterize 3D human face into identity shape $\beta\in\mathbb{R}^{|\beta|}$, facial expression $\psi\in\mathbb{R}^{|\psi|}$, and pose parameters $\theta\in\mathbb{R}^{3k+3}$ for rotations around $k$ = 4 joins (neck, jaw, and eyeballs) and global rotations. Each rotation vector of pose parameters is represented by  axis angles ($\in\mathbb{R}^{3}$). Given these parameters, FLAME outputs a 3D mesh with vertices $\mathbf{V}\in\mathbb{R}^{N_v\times3}$ and facets $\in$ $\mathbf{F}\in\mathbb{R}^{N_f\times3}$. 


\setcounter{figure}{1}
\begin{figure*}[ht]
\centering
\hspace{0cm}
\includegraphics[width=0.8\textwidth]
{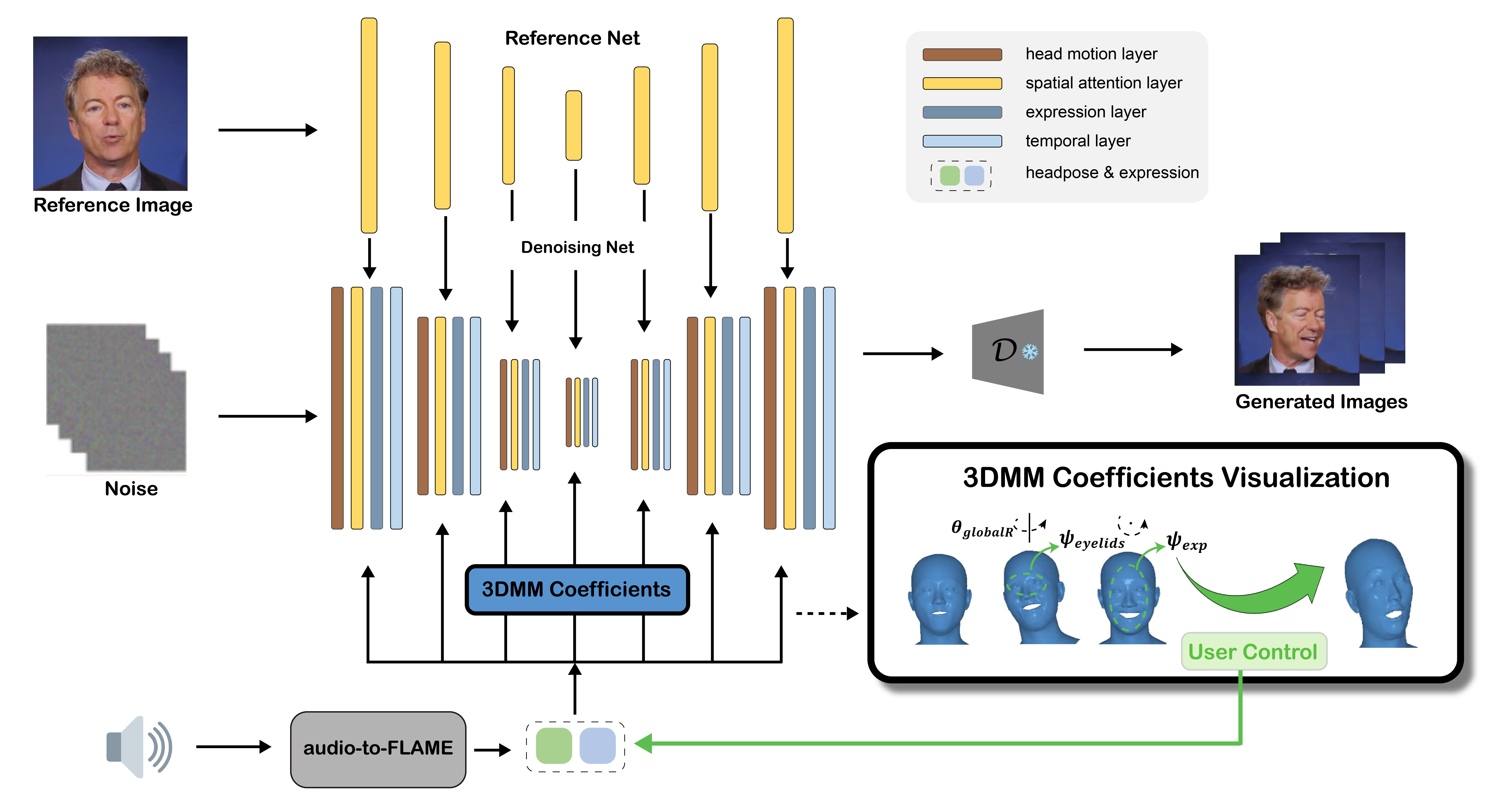}
\vspace{-10pt}
\caption{Model architecture of FLAP. The main diffusion net accept a generated 3D parameter input, which can also be modified and biased by user input. For audio only scenarios, FLAP utilizes an audio-to-FLAME module, which generate FLAME coefficients from audio. Our main U-net block consists of 4 layers, namely, head motion layer, spatial attention layer, expression layer and temporal layer. These layers are trained in different stages with different input conditions as we proposed Progressively Focused Training scheme, which will be discussed in \cref{sec:ablation}}
\label{fig:network}
\vspace{-10pt}
\end{figure*}

\vspace{-5pt}

\section{Methodology}
\subsection{Overall Framework}

FLAP is build upon Stable Diffusion (SD) and AnimateDiff \cite{guo2023animatediff} utilizing their weights for initialization. Following recent diffusion-based portrait animation methods \cite{hu2024animate}, we employed a dual U-net architecture which consists of two U-Nets initialized with SD 1.5. In this architecture, one of the U-nets, namely ReferenceNet, works as reference image encoder and, therefore, does not participate in the denoising process. The other U-net, namely DenoisingNet, is responsible for denoising process and generating videos guided by conditions provided by ReferenceNet and other modules.

As shown in \cref{fig:network} , the input of our method consists of three parts: 1) a single portrait image of an arbitrary identity, 2) a speech audio clip from an arbitrary person, and 3) optional user specified control signals. As for user control, our method support various forms of signals. For head motion control, users are allowed to use precise rotation angles around X, Y, and Z axes for each frame to generate desired head movement. Users can also provide a video or an image and extract its head pose with FLAME trackers as control signal. In addition, user can track and modify input video motions by changing its rotation angles to generate final movement. As a conclusion, FLAP only need a final angle in 6-DOF as control signal for pose, which allows users to freely overlay or modify head angles from different source to generate videos with desired head movements. Notably, if audio is provided, by overlaying mild head movements generated from audio with user-controlled movements, our method generates user-specified head actions while still preserving the subtle nodding when speaking, thus maintaining naturalness. If a fix head is needed, user can set head angle to a fixed value to generate talking head videos with fixed pose.


Intuitively, head movements and facial expressions are closely tied with the tone and cadences in speech. To ensure such an interconnection, we bind head movements, facial expressions and lip movement to audio utilizing an audio-to-FLAME module. This module operate to generate natural head movement, expressions and lip movements from audio solely, avoided expression disruption from reference image. The detail of this module will be discussed in \cref{sec:a2e}. And the visualization of this expression disruption in existing methods will be demonstrate in \cref{sec:emo of Audio-to-FLAME module}.

\vspace{-2pt}

\subsection{Controlling Diffusion Model with 3D Head Coefficient as Condition}

In recent diffusion-based works like Follow-your-Emoji \cite{ma2024follow}, V-Express \cite{wang2024v} and EchoMimic \cite{chen2024echomimic}, there has been a tendency to project landmark information into image space, and further serve as the condition to guide diffusion process. Although projected landmark images can be easily applied to control net like structures with out extra effort, it leads to a loss of head motions in 3D space. Additionally, 2D landmarks carry identity information which need careful design of alignment method across different input images. Problems are likely to arise when there is a large appearance variation in source image and driving video from which landmarks are extracted. Apart from the challenges in unifying landmarks to prevent id leakage, another problem of landmark-based approach is that there is hardly no landmark can be detected in stylized portraits, anime characters and portrait sculptures, making landmark alignment even more challenging.

To address above problems, we introduce 3D head coefficients which include head pose coefficients and facial expression coefficients to control diffusion model. In FLAP, our 3D head coefficients are based on FLAME, a parametric 3D head model. We use global rotation $\theta_{globalR}\in\mathbb{R}^{3}$ in \cref{eq:FLAME formulation} as an additional diffusion condition for head control. $\theta_{globalR}$ is a sub-vector of  $\theta$, which denotes global head rotation in 6 DoF and is represented by 3 angles of rotation around $x$, $y$ and $z$ axis. For facial expression coefficients, we use $\theta_{jaw}\in\mathbb{R}^{3}$, $\theta_{eyes}\in\mathbb{R}^{6}$, $\psi_{eyelids}\in\mathbb{R}^{2}$ and $\psi_{exp}\in\mathbb{R}^{100}$ in \cref{eq:FLAME formulation} for jaw movement, eye features, eyelids features and expressions. Thus, the proposed 3D head conditions for diffusion process in FLAP can be denotes as 
\vspace{-5pt}
\begin{equation}
c_{3Dhead}=[\theta_{globalR},\theta_{eyes},\theta_{jaw},\psi_{eyelids},\psi_{exp}]
 \label{eq:3D head conditions}
\end{equation}
Overall conditions for diffusion process can be summarize as $C=[C_{3Dhead},c_{ref}]$, where $C_{3Dhead}$ is a sequence of $c_{3Dhead}$ that contains 3D head information varying over time, with each $c_{3Dhead}$ corresponding to a frame in the video. We express this formally as $C_{3Dhead}=\{c_{3Dhead_1},c_{3Dhead_2},c_{3Dhead_3},...,c_{3Dhead_n}\}$. For all coefficients mentioned above, we follow the settings of FLAME to open up opportunities to incorporate FLAP with other models in FLAME's ecosystem such as EMOCA \cite{danvevcek2022emoca}, EMOTE \cite{danvevcek2023emotional}, and other works in upstream tasks. 

The purpose of our DenoisingNet is to generate a video that both matches the 3D head information in $C_{3Dhead}$ and the appearance information in $c_{ref}$. Thus, during training, we extract $c_{3Dhead}$ for every frame in training video using FLAME fitting method \cite{zielonka2022towards}. Then train the model to generate video frames given 3D head condition sequence $C_{3Dhead}$ of target frames along with a reference image.Target video is used as supervision. The training objective can be formulated as follow:
\vspace{-4pt}
\begin{equation}
L=\mathbb{E}_{z_t,c,\epsilon\sim\mathcal{N}(0,1),t}\left[\|\epsilon-\epsilon_\theta(z_t,t,C)\|_2^2\right]
  \label{eq:training objective}
\end{equation}
where $\epsilon_\theta$ denotes DenoisingNet. During inference, as shown in \cref{fig:network}, a given portrait image is feed into ReferenceNet to get $c_{ref}$. And a sequence of 3D head condition $\{c_{3Dhead_1},c_{3Dhead_2},c_{3Dhead_3},...,c_{3Dhead_n}\}$ is generated by audio-to-FLAME module and then manipulated by user signals, and then feed into DenoisingNet to output a talking head video precisely controlled by 3D head conditions. 

\vspace{-2pt}

\begin{table*}[h!]
  \begin{center}
    \caption{Quantitative comparisons with existing works. 1\stars{1}in controlability means only support audio or video driven. 2\stars{1}indecates support both ways. 3\stars{1} means head pose and facial expression can be controlled independently. FLAP has 4\stars{1}since it not only supports all features above, but also supports angle input for nuance head pose manipulation and control.The best result for each metric is in \textbf{bold}, and the second-best result is \underline{underlined}.}
    \label{tab:exp_result}
    \vskip 0.1in
    \begin{tabular}{l|l|l|c|c|c|c|c}
      \hline
      \textbf{Model} & \textbf{Controllability} &\textbf{Driven}& \textbf{FID}$\downarrow$& \textbf{FVD}$\downarrow$& \textbf{FaceIQA}$\uparrow$& \textbf{SYNC-C}$\uparrow$& \textbf{H-IQA}$\uparrow$\\
      \hline
      MCNet& \stars{1}  &video& -& -& 0.4001& 7.4600 &0.2992\\
      EchoMimic& \stars{2}  &video& -& -& 0.5890& 6.7896&  0.3919\\
      EDTalk& \stars{3}  &video& -& -& 0.4159& \textbf{8.4260}&  0.3018\\
      X-Portrait & \stars{1}  &video& -& -& \textbf{0.6544}& 5.8181& \textbf{0.4795}\\
      LivePortrait & \stars{1}  &video& -& -& 0.5474& 7.6228& 0.3987\\
      \hline
      \textbf{FLAP (ours)} & \stars{4}  &video& -& -& \underline{0.6107}& \underline{7.7837}&  \underline{0.4644}\\
      \hline
      SadTalker& \stars{1}  &audio& 163.31& 551.94& 0.4798& 6.9274& 0.3137\\
 EDTalk& \stars{3}  &audio& 166.11& 592.20& 0.4434& 7.2472&0.3052\\
      EchoMimic & \stars{2}  &audio& 148.91& 500.56& 0.6116& 7.0341& 0.4329\\
      AniPortrait& \stars{1}  &audio& \underline{137.27}& 581.01& \textbf{0.6518}& 7.6117& \textbf{0.4920}\\
      Hallo& \stars{1}  &audio& 139.54& \underline{457.43}& 0.610& 7.5069& 0.4417\\
 Hallo2& \stars{1}  &audio& 141.54& 483.95& 0.5993& \underline{7.7184}& 0.4241\\
        \hline
      \textbf{FLAP (ours)} & \stars{4}  &audio& \textbf{136.54}& \textbf{399.78}& \underline{0.6254}& \textbf{7.8617}& \underline{0.4592}\\
      \hline
    \end{tabular}
  \end{center}
\vskip -0.1in
\end{table*}

\subsection{Progressively Focused Training for Expression-Motion Decoupling}

Training process in previous methods \cite{tian2024emo, ma2024follow, hu2024animate, xu2024hallo, wang2024v} is structured into two stages, namely image training stage and video training stage. For image training, two frames are randomly extracted from a video: one frame is used as the reference image, while the other serves as the target. With reference image used as input for ReferenceNet, the whole model is trained to make predictions conditioned on given reference image without audio input. In video training stage, the model is trained to predict a continuous sequence of frames conditioned on both audio and reference image. However, 3D head coefficients delivers global pose information and detailed face information at the same time. Thus, when they are introduced as condition, model struggles to learn a correct patterns from massive volume of information within one stage.

Additionally, another significant challenge is learning a decoupled representation of facial expression and head pose. Although FLAME model itself decoupled facial expression from head pose, an unavoidable nature of monocular face 3D reconstruction is that the head has one side facing the camera while the other side is not. Thus, when using popular fitting methods \cite{zielonka2022towards, retsinas20243d}to reconstruct 3D head coefficients from monocular videos, it's inevitable that head pose information will leak from expression coefficients. Given a video with only the left side of a talking face, 3D head fitting methods can not get correct coefficients of the other side of this face, leading to inaccuracies and flaws in the representation of the unseen half of face. Therefore, a pattern emerges that if certain facial coefficients appear anomalous, it indicates that this side of the head is not facing the camera. As a result, expression coefficients contain a learnable pattern of head pose information. This makes it easy to train a biased model. Visual results are provided in \cref{sec:ablation}.

To address above issues, we proposed a Progressively Focused Training strategy (PFT) for training efficiency and expression-motion decoupling. Guided by a basic concept that models tend to learn global features of an image before focusing on the finer details. We divide 3D head conditions in FLAP into two groups: detailed expression conditions $c_{exp}=[\theta_{eyes},\theta_{jaw},\psi_{eyelids},\psi_{exp}]$, and head motion condition $c_{motion}=[\theta_{globalR}]$. Head motion condition is learned in earlier stage, while expression conditions is learned in later stages. Thus, our progressively focused training strategy separate training process into three stages. The first stage is head motion stage, which we conduct image training with all layers trainable. In this stage, $c=[c_{head},c_{ref}]$ is the only provided condition. The model is forced to focus on head pose changes between reference image and target image. The second stage is expression stage. In this stage, the model focus on learning facial expressions which includes eyes information $\theta_{eyes}$ and $\psi_{eyelids}$, jaw information $\theta_{jaw}$ and global expression information $\psi_{exp}$. The overall conditions in this stage can be formulated as $c=[c_{exp},c_{ref}]$, where $c_{exp}=[\theta_{eyes},\theta_{jaw},\psi_{eyelids},\psi_{exp}]$. To avoid coupling of expression and head motion, attention layers related to head motion are frozen. In the third stage, video training is conducted, we follow the settings in recent works \cite{tian2024emo,xu2024hallo} to train Animatediff \cite{guo2023animatediff} blocks.

\vspace{-2pt}

\subsection{Audio to FLAME Module}
\label{sec:a2e}

Most existing diffusion methods trends to generate expressions and appearance jointly in a DenoisingNet with condition $c=[c_{audio},c_{ref}]$. We argue that in training, the existence of $c_{ref}$ can disrupt the learning of $c_{audio}$. From our observation, reference images carry more learnable information of expressions than audio. Since the speaker's emotion often show minimal variation in short videos. Therefore, the model will be misled to learn a biased pattern by simply copy the expressions of reference image to output video. For example, given a smiling face image and angry speech, model might generate a half-smiling video, fail to align with the angry audio input. More visual results of this observation can be found in \cref{sec:emo of Audio-to-FLAME module}.

To avoid $c_{audio}$ being affected by $c_{ref}$, we ultilize an audio-to-FLAME module which generate FLAME-based expression coefficients and head motion coefficients solely from audio. This helps us bind facial expressions and head motion to audio, eliminating $c_{ref}$'s impact on expressions. Another advantage of this is that user can control pose and expression freely by manipulating FLAME coefficients. Benefiting from this architecture, users can also extract coefficients from other videos to accomplish video driven portrait animation tasks. An interactive model viewer\footnote{ https://flame.is.tue.mpg.de/interactivemodelviewer.html } provided by the authors of FLAME visualize the mapping of FLAME parameters to the mesh output. 

Generating head pose and expression from audio in 3DMM coefficients space is a well studied task \cite{sun2024diffposetalk,peng2023emotalk}. Our Audio-to-FLAME module follows the designed of \cite{wen2020photorealistic}. We retained the network with FLAME coefficients instead of BFM. As FLAME coefficients is portable, the audio-to-Flame module can be replaced by any FLAME coefficients generating methods such as EmoTalk\cite{peng2023emotalk},EMOTE \cite{danvevcek2023emotional}, DiffPoseTalk\cite{sun2024diffposetalk}. 

\begin{figure*}[ht]
\centering
\includegraphics[width=.9\textwidth]
{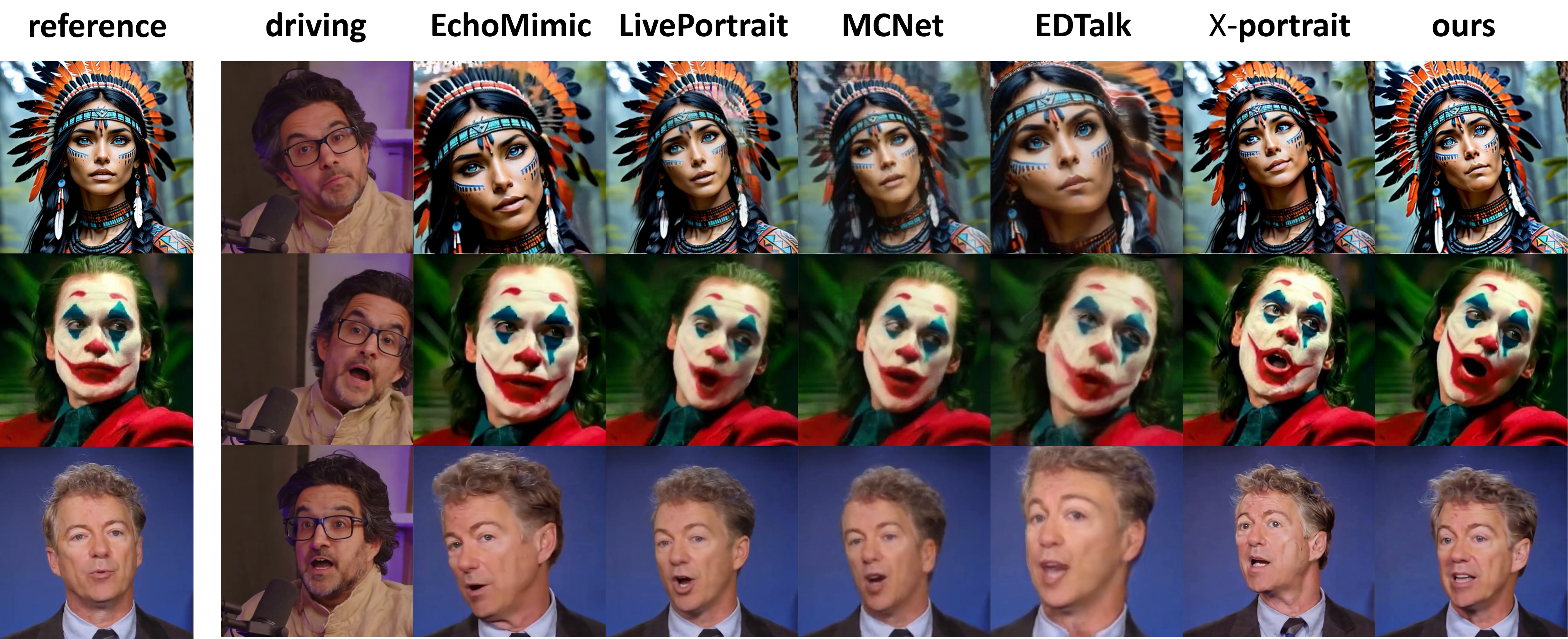}
\vspace{-5pt}
\caption{Qualitative comparisons. Our model achieves the most accurate pose control, visual quality and lip synchronization.}
\label{fig:vis_cmp}
\end{figure*}

\begin{figure*}[ht] 
\centering
\includegraphics[width=.9\textwidth]
{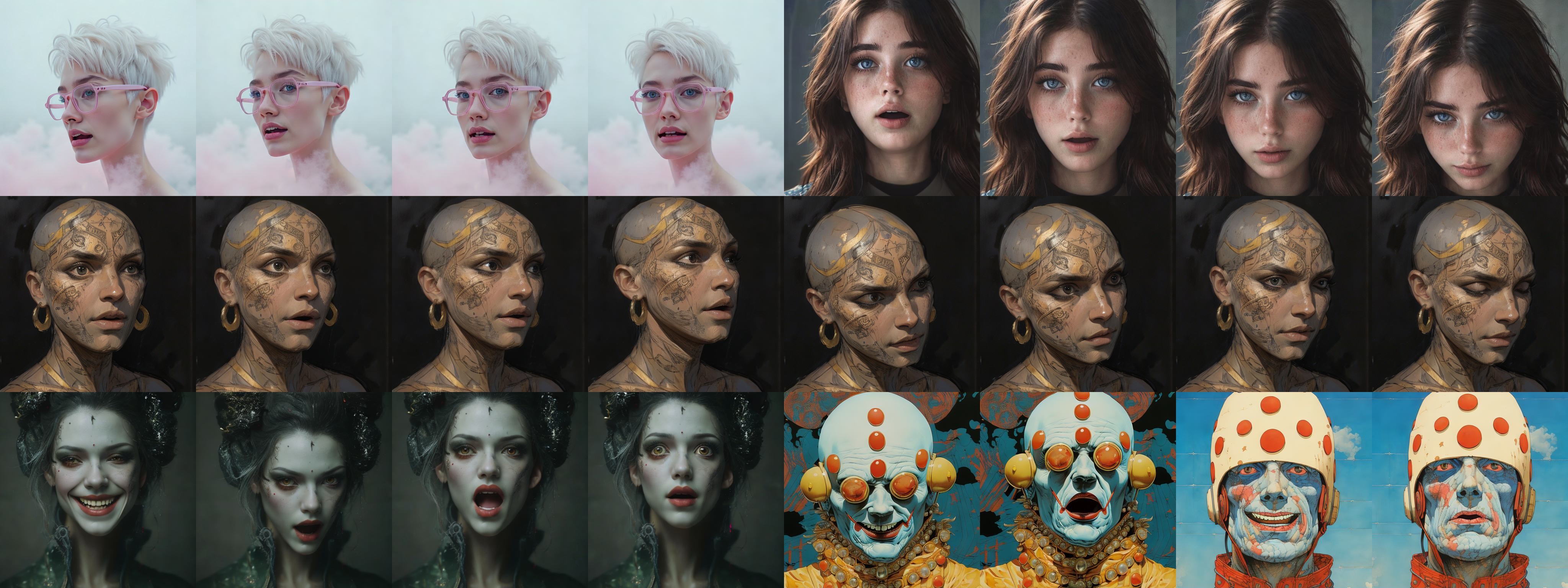}
\vspace{-10pt}
\caption{Visual Results on 3 dof head pose control and rich expressions. Expressions of reference images are all neutral}
\label{fig:pose_exp}
\vspace{-10pt}
\end{figure*}

\section{Experiments}
\subsection{Implement Details}
Our model is trained on HDTF, Nersemble, Celebv-HQ, VFHQ, Celebv-text and data collected from youtube. All video frames are resized to 512$\times$512. For face reconstruction and tracking, we re-implemented MICA \cite{zielonka2022towards} for video tracking as our tracker to fit FLAME coefficients. The dimension of 3D head condition $\theta_{globalR}$, $\theta_{eyes}$, $\theta_{jaw}$, $\psi_{eyelids}$, and $\psi_{exp}$ is 3, 12, 3, 2, and 100, respectively. In head motion training stage, we use videos with head motion coefficient variance in top 20\% and train for 150,000 steps. For facial expression training and video training, we train our model with all data for 150,000 and 50000 steps.

\subsection{Visual Results on Controllable Generation}
\cref{fig:pose_exp} presents some results generated by our method. Given a sequence of head angles as control signal, our model can generate film-like scene with an avatar performing user designed actions. The first row demonstrate two avatars turning his/her head around and making a head-down gesture. The second row shows an avatar first looking around and then performing another sequence of user defined actions. In the third row, we demonstrate the model's ability of generating life-like expressions for stylized characters and novel anime characters, despite our model being trained exclusively on real portrait videos. With these abilities, our model breaks down the barriers in filmmaking, enabling real contributions to the creative process.

\subsection{Evaluation}
\subsubsection{Quantitative Evaluation}
As evidenced in our qualitative comparisons in \cref{tab:exp_result}, FLAP outperforms other methods on most of the metrics in audio driven task. We achieve the best FID, FVD, SYNC score and second best IQA. Among these baselines, our method offers the best controllability without compromising its audio driven functionality. For IQA scores, Aniportrait achieves the highest scores. However, we argue that Aniportrait tend to generate videos with less expression and head movement so that there is less artifacts. We do a more detailed comparison with it in \cref{sec:anip}.

For video driven task, shown in \cref{tab:exp_result}, we do a cross-identity reenactment. Although EDtalk has the best Sync scores, its poor H-IQA and Face-IQA indicates its shortcoming in visual quality. X-portrait achieves the best visual quality, but has the worst lip synchronization among all methods. Our model achieve the second best in all metrics, successfully balanced visual quality and lip synchronization. It should be highlighted that our method only use a 120-dim FLAME coefficients vector for driving, while other methods rely on the entire video as their driving signal. 

\subsubsection{Qualitative Evaluation}
For qualitative comparisons, as shown in \cref{fig:vis_cmp}, our method generates the best results in common scenarios and scenarios with wide-range of head rotations, complex headpieces and makeup. Compared to our methods, Echomimic uses projected landmark images as control signal for generation and yields animations results with inaccurate pose and artifacts in headpieces. We perceive this inaccuracy as the loss of 3D pose information when projecting 3D key points to 2D images. As for X-portrait, it generate texture with high quality which results in better IQA scores, but it suffers from severe distortion around face and neck. Among all these methods, our method generate portrait videos with superior perceptual quality, motion precision and less artifact in both common scenarios and more complex scenarios.

\begin{figure}
\centering
\includegraphics[width=1.0\linewidth]
{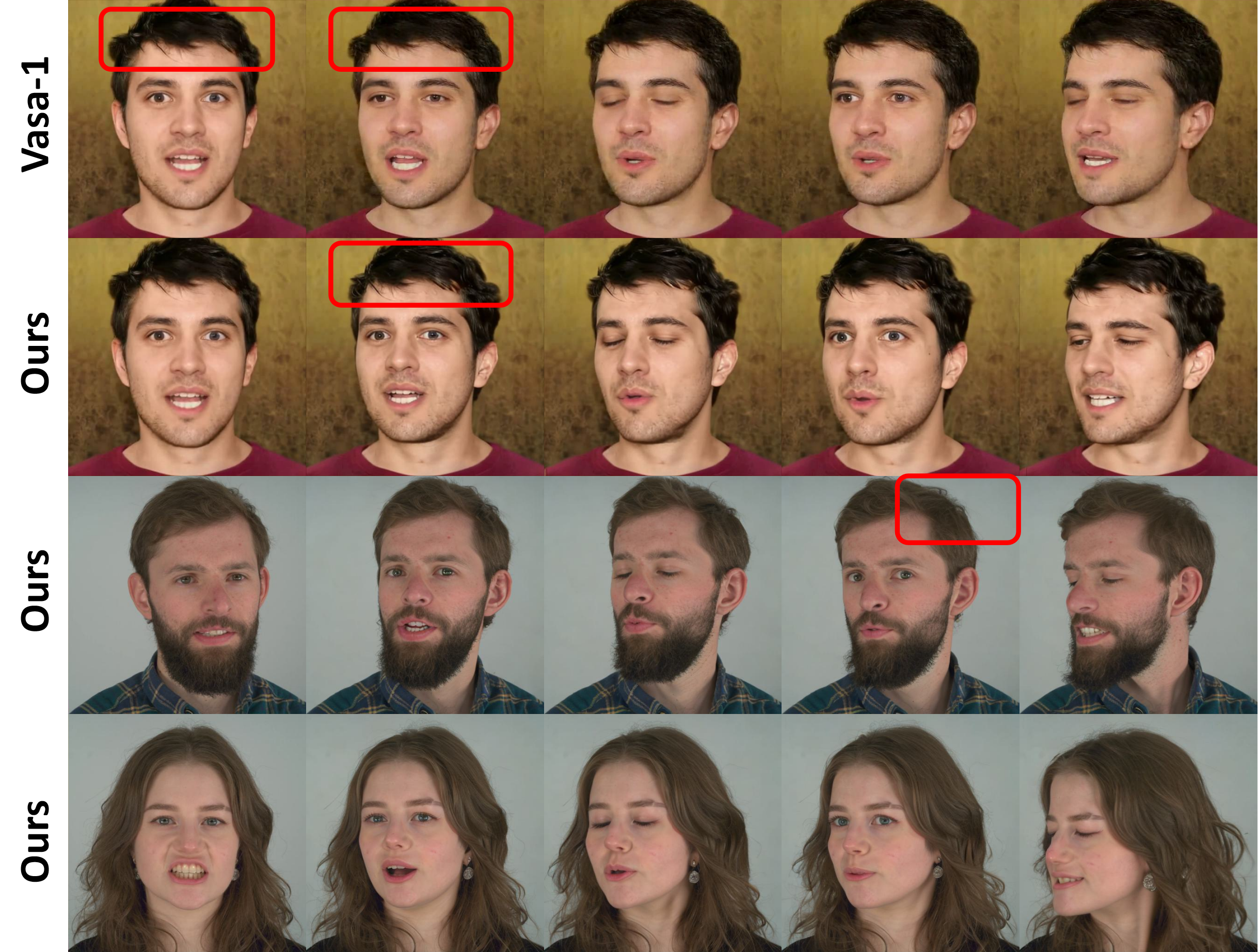}
\vspace{-15pt}
\caption{Comparison with VASA-1. The first column shows the reference image, the rest of the column shows generated results. The captions above shows the hair feature is gradually missing in VASA-1 demo video.}
\label{fig:vasa-cmp}
\vspace{-10pt}
\end{figure}
\vspace{-5pt}

\subsection{Comparison with VASA-1}
Since VASA-1 is not an open source project, we can only provide a qualitative comparison with its demo from the official website. As shown in \cref{fig:vasa-cmp}, the first column demonstrates the original reference image, with his distinguishing hair feature highlighted in red box.  We find that although VASA-1 achieve accurate pose control with user pre-defined angular information as ours method does, it suffers from the mean face problem as most GAN based decoders do. While the angle increase, the character starts losing its hair and face features and finally differs from original reference image. Vasa model seems tend to generate a face of an average man and fail to capture the detailed side profile feature of reference image. From
our observation, this frequently happened in FaceVid2Vid\cite{wang2021one} and Mega-Portrait. Vasa uses a retrained Mega-Portrait as its image encoder and decoder, which limits its expressiveness and causes this mean face problem. The third and the last rows shows that our method successfully avoid this mean face problem and generate detailed profile feature across different identities.

\begin{figure}
\centering
\includegraphics[width=1\linewidth]
{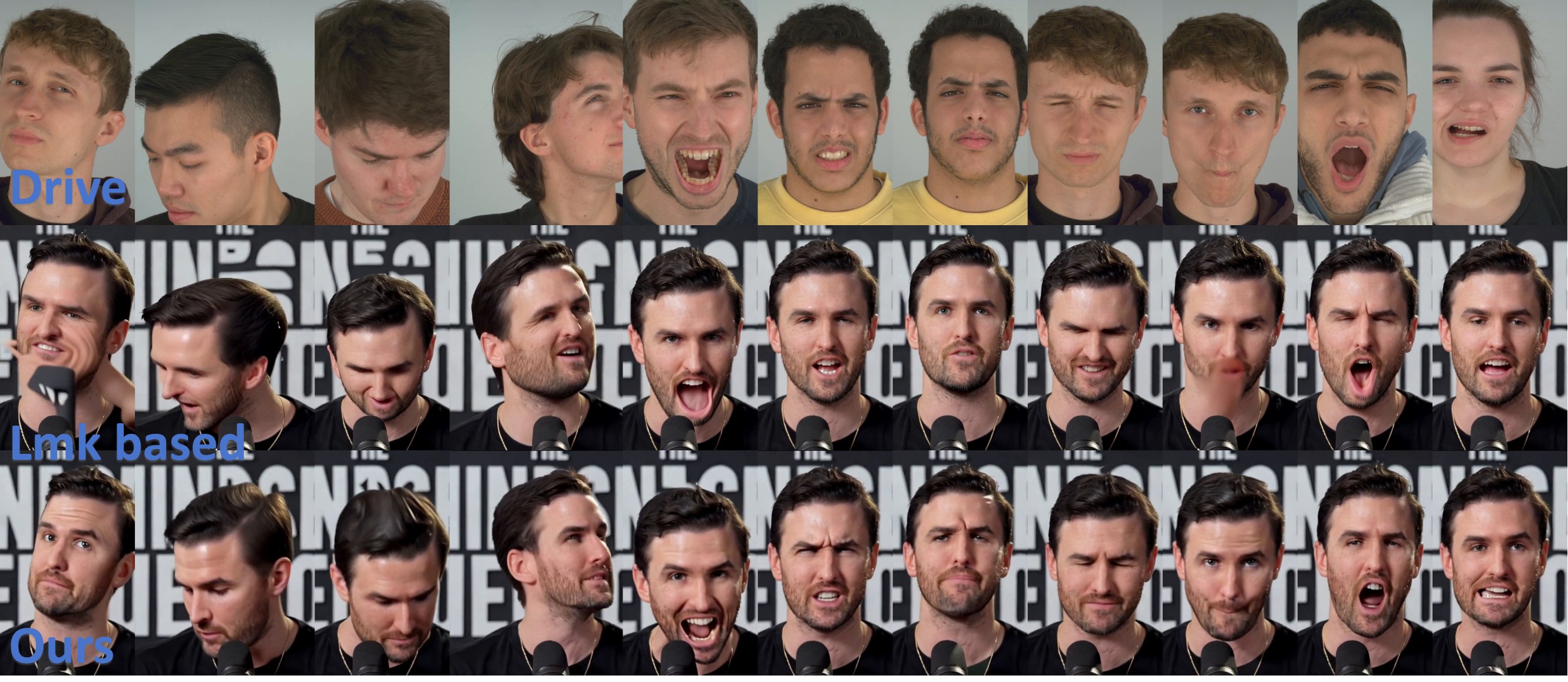}
\vspace{-10pt}
\caption{Comparison with Landmark-based methods. The cropped images are only used for demonstration, we keep the all input dimensions as 512x512}
\label{fig:anip-cmp}
\vspace{-5pt}
\end{figure}

\subsection{Comparison with Landmark-based methods}
\label{sec:anip}
Prior studies like Aniportrait and Echomimic use facial landmarks as guidance for controllable generation. Here we provide a more detailed comparison between our method and Aniportrait. For Echomimic, visual comparison can be found in \cref{fig:vis_cmp}. As shown in \cref{fig:anip-cmp}, Lmk-based methods refers to Aniportrait. It is not capable of generating head movement like spinning and nodding since the projected 2D landmarks have lost most of the 3D pose information, thus offering vague pose information in denoising process. In \cref{fig:anip-cmp}, the head generated by Aniportait in the first 3 columns from left are all squished due to this projection error. In the remaining columns, aniportrait failed to generate accurate facial dynamics when there are significant differences between the driving identity and reference identity. As we disscussed in former sections, these are the natural shortcomings of landmarks serving as control signals. However, in audio-driven mode, Aniportrait hardly generate head movement or dramatic facial expressions from audio, making it easier to maintain visual quality than other methods. On the other hand, our model use unified FLAME coefficients as conditions, avoiding the need for alignment and thus generating accurate facial dynamics and head pose as shown in \cref{fig:anip-cmp}. Our approach demonstrates that FLAME coefficients can capture more subtle expressions, and is a more suitable as intermediate representation for controllable generation of talking head avatars.

\subsection{Analysis and Ablation Study}

\begin{figure}
\centering
\includegraphics[width=1\linewidth]
{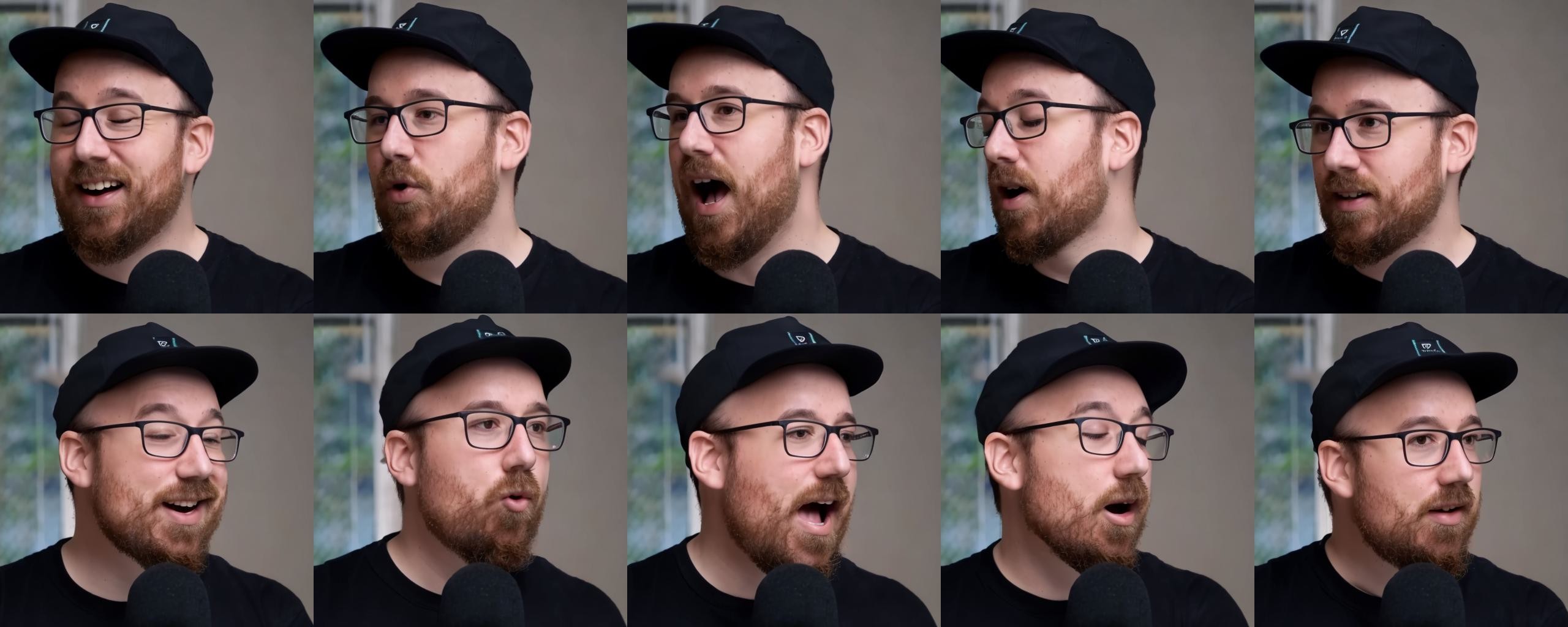}
\vspace{-20pt}
\caption{Visual results of pose editing of an speech video. The first row shows the original video. We edit his head pose toward right in the second row.}
\label{fig:user-ctrl}
\vspace{-10pt}
\end{figure}

\subsubsection{Analysis on Pose Editing}
An example of our fine-grained control of expression and pose is demonstrated in Figure 1. Here we further demonstrate the visual results of pose editing of a speech video clip. In \cref{fig:user-ctrl}, we edit the the orientation of the person's head in the video clip without effecting his lip movement and expression (first row to second row). Our method exhibits the capability to disentangle pose and facial dynamics.

\begin{figure}
\centering
\includegraphics[width=1\linewidth]
{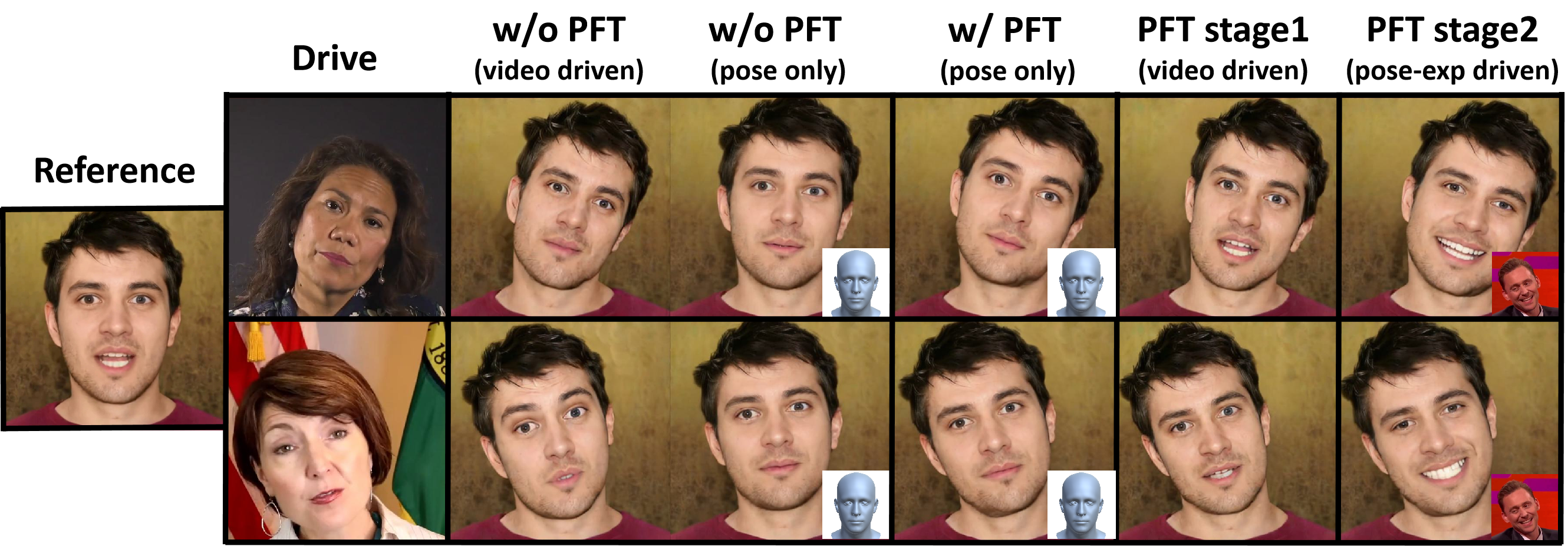}
\vspace{-15pt}
\caption{Ablation study on Progressive Focused Training (PFT). Pose only indicates that we use the expression coefficients from a canonical head who has a neutral face and canonical head pose (visualized at the bottom right of related images). By swapping original expression with this idle expression, we can explore whether the output head pose is affected by expression coefficients. And thus explore the head pose information leakage from expression coefficients. In the last column, the pose is driven by the 2nd column, while the expression is driven by the bottom right sub-image.}
\label{fig:ablation study}
\vspace{-15pt}
\end{figure}

\subsubsection{Ablation Study}
\label{sec:ablation}
We conduct an ablation study to demonstrate the effectiveness of our proposed Progressively Focused Training (PFT) scheme. As is shown in the \cref{fig:ablation study} (the 2nd and 3rd left column), although FLAP can be trained to fit the source pose and expression well on video driven mode without PFT scheme, its head pose is heavily relied on signals leaked from expression coefficients, indicating failure in decoupling. To confirm this assumption, in the 4rd column, we set its expression coefficients idle, and get a result with decrease of head movement range. By contrast, in 5th column, when PFT is applied, the expression change will not affect head pose and achieve full disentanglement. In addition, when the model only completes the head motion training, it can only generate head motion aligning with pose source, but is not capable of modifying the expression of reference image (6th column). In the last column, when the model completes expression training stage, it's capable of generating both head pose and expression from two different source. This decoupling of head pose and facial expression allows for independent control and shows the effectiveness of proposed PFT. Moreover, the proposed PFT scheme can be generalized to other conditions during training. We test this scheme useing feature vectors from EDTalker \cite{tan2025edtalk} in supplementary.

\begin{figure}
\centering
\includegraphics[width=.86\linewidth]
{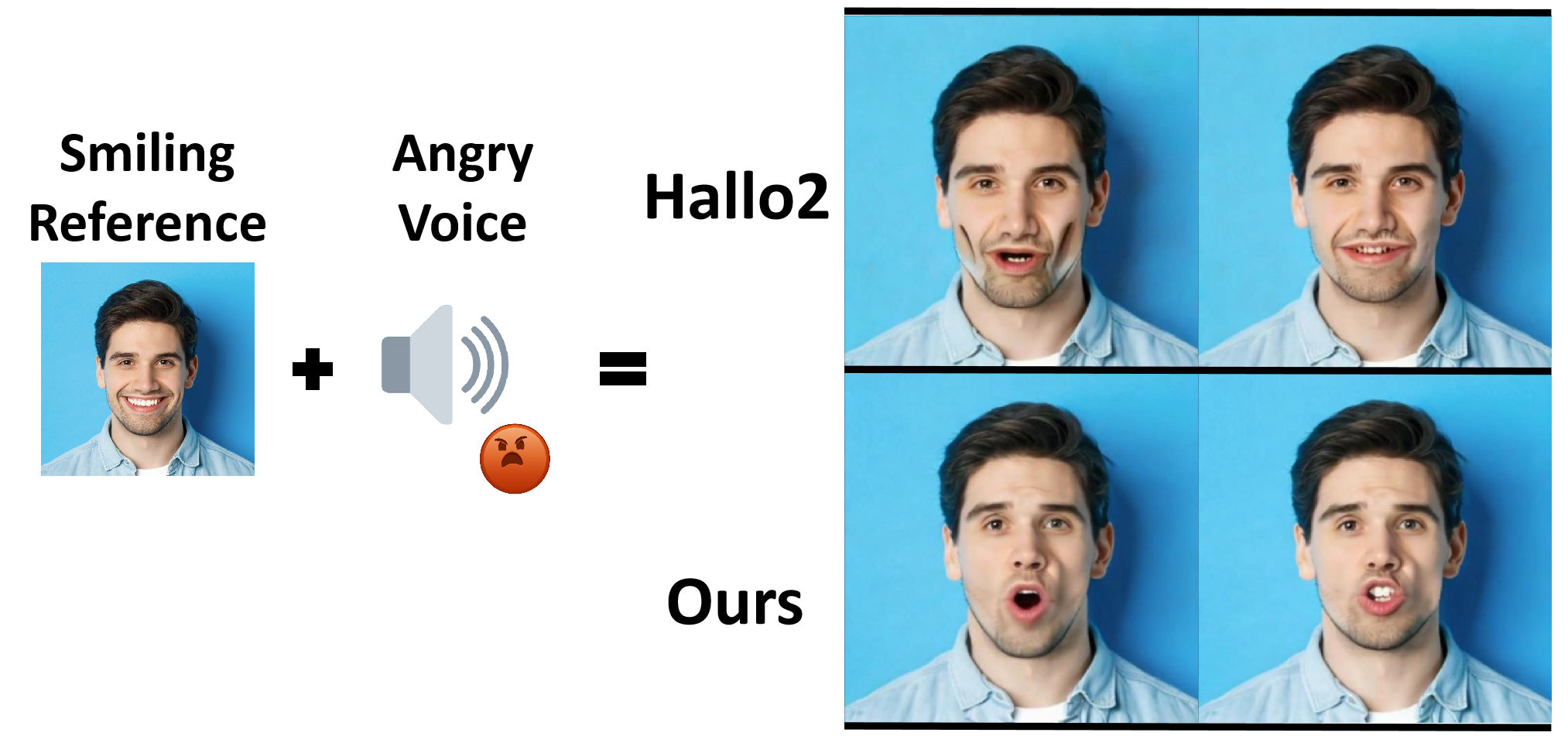}
\vspace{-10pt}
\caption{Visual result of our method and Hallo2 \cite{cui2024hallo2} animating a smiling portrait with angry tone.}
\label{fig:bias}
\vspace{-20pt}
\end{figure}

\subsection{Effectiveness of Audio-to-FLAME module}
\label{sec:emo of Audio-to-FLAME module}
Our Audio-to-FLAME module is capable of binding the facial expression to audio, making all expressions irrelevant to reference image. As shown in \cref{fig:bias}, Hallo2 struggles to generate appropriate expressions since it is confused by the smiling face in reference image and angry tone in the audio. Therefore, it generates a mixed and unnatural face. On the contrary, our model generate expressions solely relies on audio though our Audio-to-FLAME module, and thus generate accurate and natural expressions.

\section{Conclusion}
In this paper, we present FLAP, a novel approach for zero-shot talking-head avatar generation that integrates explicit 3DMM coefficients into a diffusion-based framework. By enabling independent control over head movements and facial expressions with PFT, FLAP significantly improves both naturalness and controllability of generated portrait videos. Our model demonstrates a nuanced understanding of 3D space, offering flexibility in controlling head movements and generating lifelike expressions from diverse sources. Extensive experiments show that FLAP outperforms existing methods in terms of visual quality lip synchronization and controllability, providing a promising solution for real-world applications. 


\setcounter{figure}{0} 
\begin{figure*}[!h]
\centering
\includegraphics[width=.95\textwidth]
{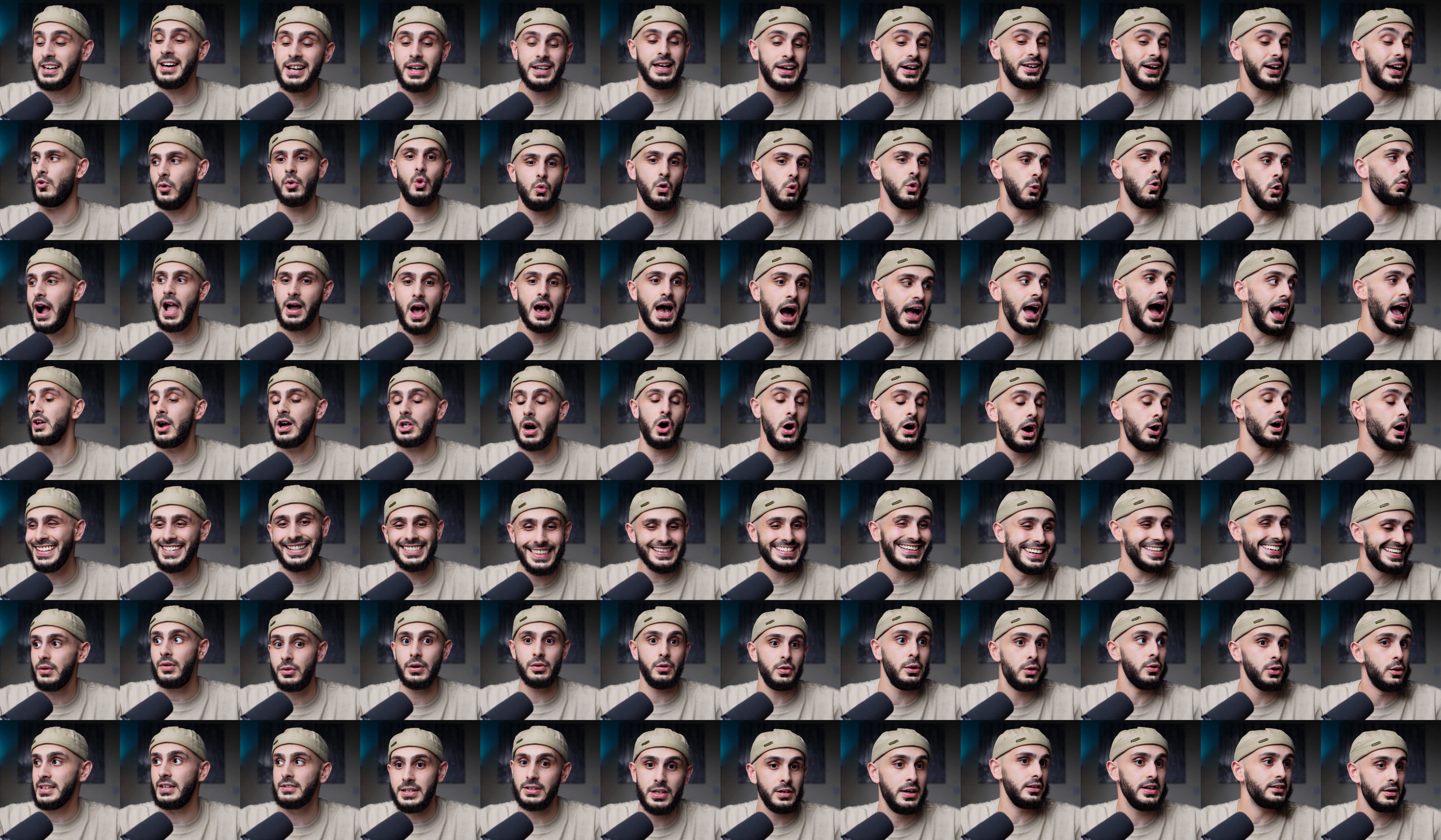}
\vspace{-10pt}
\caption{Additional Results on Multi-pose Generation}
\label{fig:multi_pose}
\end{figure*}
\vspace{-10pt}

\renewcommand{\thesection}{\Alph{section}}
\setcounter{section}{0} 
\section{Appendix}

\subsection{Additional Results on Multi Pose Generation}
\cref{fig:multi_pose} provides more results of multi-pose and multi-expression generation of the same identity. The range of head rotation angle is pre-defined by a 3-dim vector representing [x,y,z] angle in 3 DoF as control signal from user.

\subsection{Flexibility Analysis}
\subsubsection{Portability of FLAME Coefficients}
\begin{figure}[!h]
\centering
\includegraphics[width=1.\linewidth]
{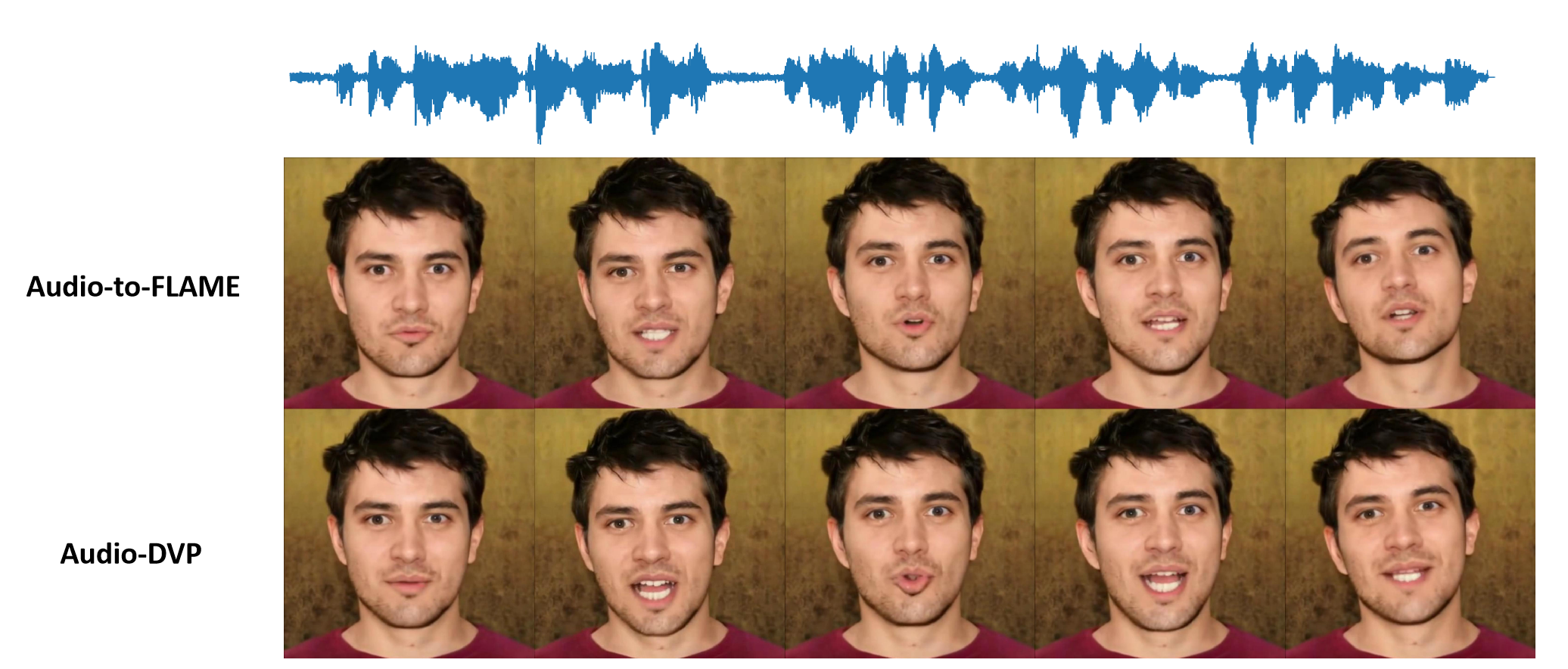}
\vspace{-15pt}
\caption{Visual result of integrating talking style from \cite{zhang2023sadtalker} into Audio-DVP\cite{wen2020photorealistic} and cooperate with FLAP. The first row demonstrate original FLAP, while the second row shows the result of utilizing Audio-DVP + style control as a replacement of Audio-to-FLAME module.}
\label{fig:audio-dvp}
\vspace{-5pt}
\end{figure}

As discussed in Sec4.4, FLAME coefficients is portable, making our audio-to-Flame module replaceable by any FLAME coefficients generating methods such as EmoTalk \cite{peng2023emotalk}, EMOTE \cite{danvevcek2023emotional}, DiffPoseTalk \cite{sun2024diffposetalk}. For demonstration, we implemented a talking style controllable FLAME talker by integrating talking style control mechanism from \cite{zhang2023sadtalker} into Audio-DVP \cite{wen2020photorealistic}. \cref{fig:audio-dvp} demonstrates the visual results of integrating this style controlled Audio-DVP into FLAP. As shown in \cref{fig:audio-dvp}, with this audio-dvp model, the style of mouth while speaking is changed, compared with original Audio-2-FLAME based FLAP, a greater degree of jaw opening is generated making it easier for lip reading. Thus, by combining with various existing methods, we can generate a range of speaking styles that we desire.

\subsubsection{Using Feature Vectors as Conditions}
\begin{figure}
\centering
\includegraphics[width=1.\linewidth]
{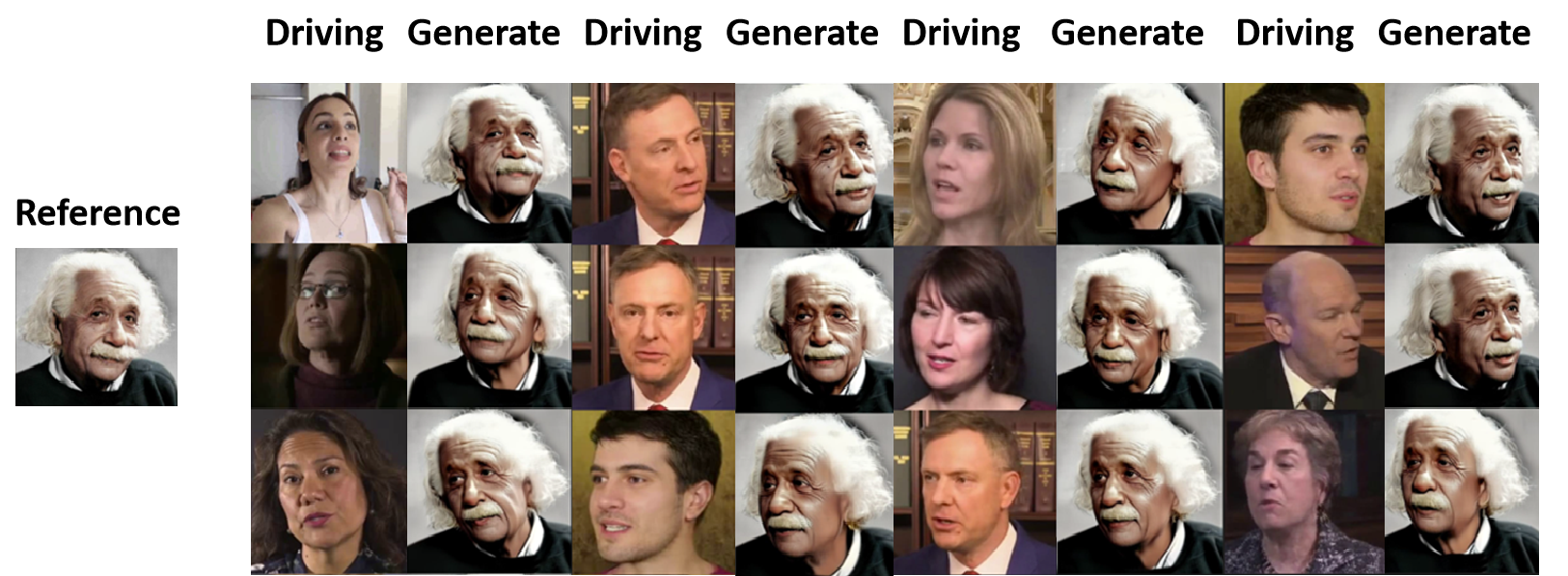}
\vspace{-10pt}
\caption{Visual results of FLAP framework cooperating with PD-FGC \cite{wang2023progressive} by retraining FLAP using feature vectors from PD-FGC as conditions. We perform pose alignment of driving image using pose vector from PD-FGC.}
\label{fig:pd-fgc}
\vspace{-5pt}
\end{figure}

\begin{figure}
\centering
\includegraphics[width=1.\linewidth]
{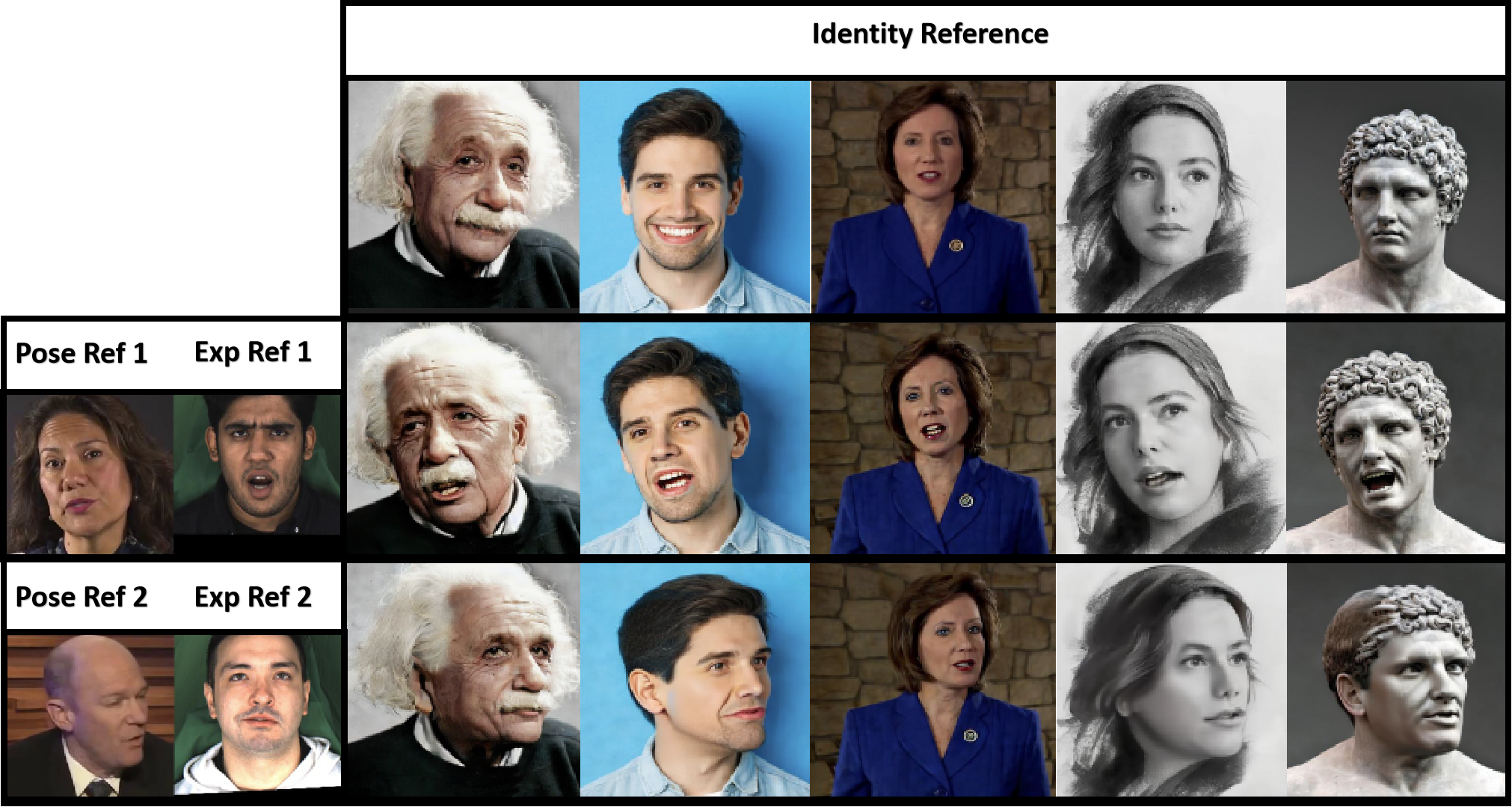}
\vspace{-10pt}
\caption{Visual results of FLAP framework cooperating with EDtalk \cite{tan2025edtalk}. The first and second columns are head pose and expression references respectively. The first row shows the id reference. The others are generated results with mixed pose, expression and id of corresponding reference.}
\label{fig:edtalk}
\vspace{-5pt}
\end{figure}

FLAP exhibits a high level of flexibility as it can be integrated with existing models by using their feature vectors as diffusion condition. For example, EDtalk \cite{tan2025edtalk} and PD-FGC \cite{wang2023progressive} proposed methods to extract implicit representations of facial expressions, head pose, lip pose and identity information from a single portrait image. Therefore, we can replace our 3D head head conditions in FLAP with feature vectors provided by EDtalk or PD-FGC. \cref{fig:pd-fgc} demonstrates the results of pose alignment using pose vector from PD-FGC. \cref{fig:edtalk} shows the results of independent control of pose and expression using pose and expression feature vector from EDTalk. Although we achieve controllable using these feature vectors, its only yield results with less naturalness and expressiveness than our original model.

\subsection{Additional Effectiveness Analysis of PFT Scheme}
To test the generalize ability of our Progressively Focused Training (PFT) scheme, we retrained our FLAP frame work using feature vectors from EDtalk \cite{tan2025edtalk} as conditions for denoising. From our observation, the feature vector of EDtalk is not fully disentangled since its generator is trained to extract desired expression and pose information from these feature. Thus, PFT is necessary in training process to make sure disentanglement of pose and expressions. As shown in \cref{fig:edtalk}, this retrained FLAP generates results with their pose aligned with pose reference image and expression aligned with expression reference image shown in the first two columns. This not only shows the generalize ability of PFT, but also demonstrate the flexibility of our model structure.

\begin{figure}[htbp]
    \centering
    \begin{minipage}{0.45\linewidth}
        \centering
        \includegraphics[width=0.85\linewidth]
        {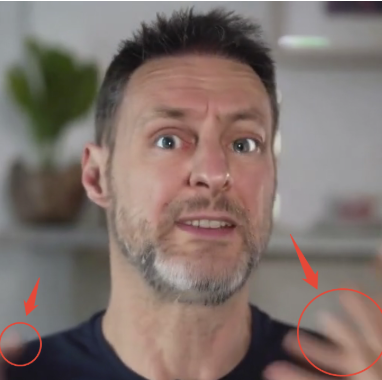}
        \caption{Hand artifact of existing models.}
        \label{fig:hand}
    \end{minipage}
    \hfill
    \begin{minipage}{0.45\linewidth}
        \centering
        \includegraphics[width=0.85\linewidth]
        {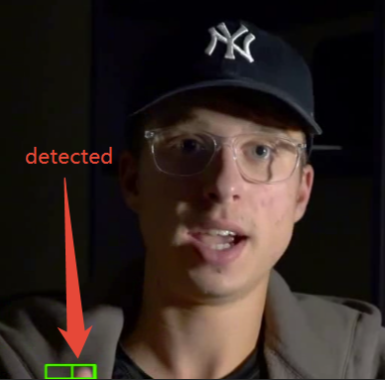}
        \caption{Detection of a small part of the fingertip.}
        \label{fig:gdino}
    \end{minipage}
\end{figure}



\subsection{Hands Filtering in Data Pre-Processing}
Filtering frames with hands in training data is crucial for the final generation results. As shown in \cref{fig:hand}, existing open sourced models like Echomimic and Aniportrait is trained on unclean dataset in which hands often appears. To train a model that will not generate artifact of hands or fingers, we utilize Grounding Dino to detect hands and fingers every 3 frames in a video. As shown in \cref{fig:gdino}, even a little spot of finger can be detected and cut out from the video. Thus, we successully trained a clean model on clean dataset.

\subsection{More Animation Results}
\begin{figure*}[!h]
\centering
\includegraphics[width=.85\textwidth]
{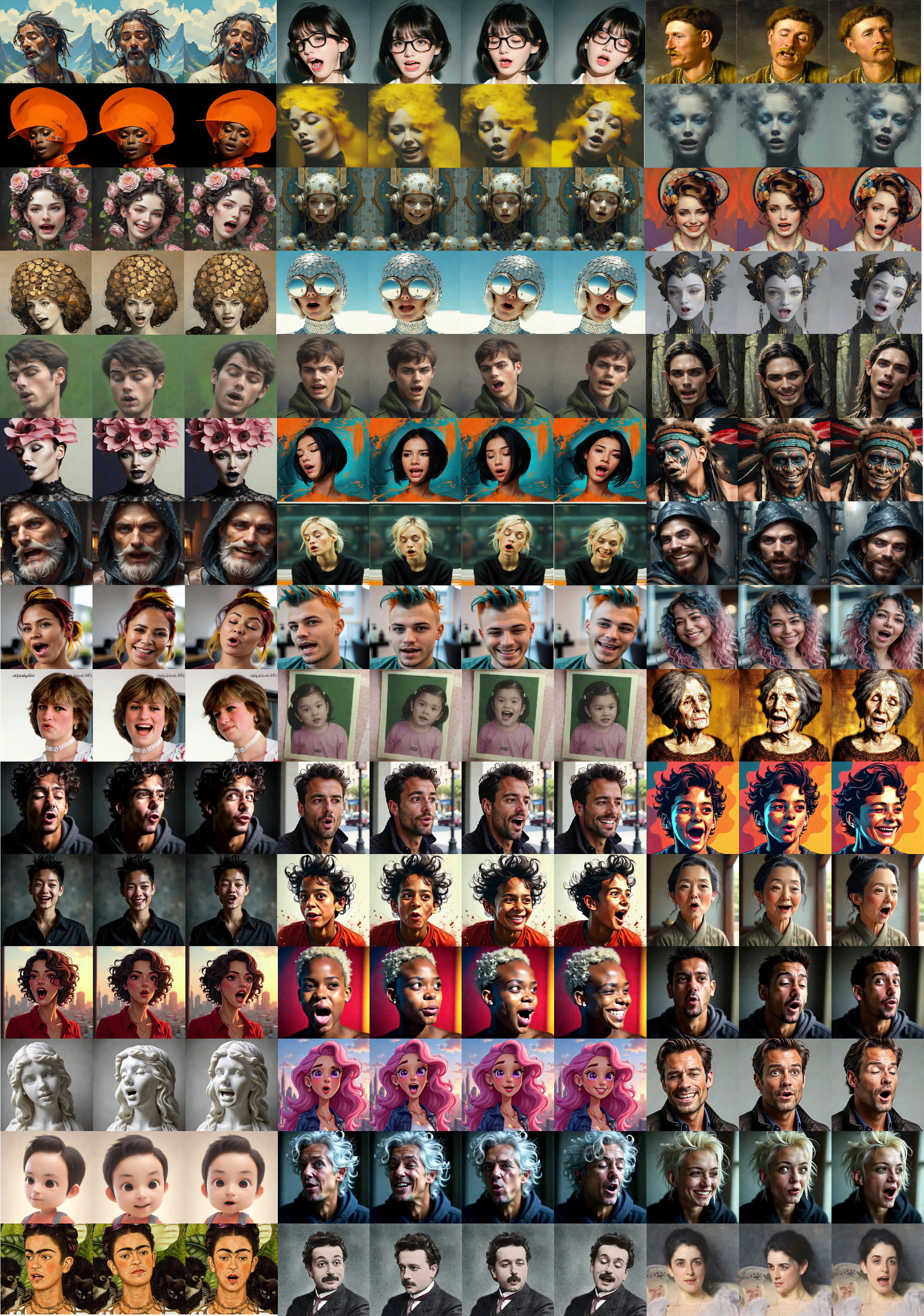}
\caption{More Animation Results}
\label{fig:demo_all}
\end{figure*}

We provide more animation results in the next page, to demonstrate the generalization ability our our model and also to illustrate their visual quality. Please refer to \cref{fig:demo_all} in the \textbf{NEXT PAGE} to see diverse expressions, poses and styles of our generated results.

\newpage

\nocite{langley00}

\bibliography{example_paper}
\bibliographystyle{icml2025}

\end{document}